\documentclass{svproc}
\usepackage{url}
\usepackage{makeidx}
\usepackage{graphicx} 
\usepackage{newtxmath}
\usepackage[bottom]{footmisc}
\usepackage{newtxtext} 
\usepackage{caption}
\usepackage{subcaption}
\usepackage{graphicx}
\usepackage{cite}
\usepackage[colorlinks=true, allcolors=blue]{hyperref}

\begin{document}
	\mainmatter        
	\title{ Analysis of Solitons within the framework of the fractional Zakharov–Kuznetsov equation utilizing  Hirota bilinear method}
	\titlerunning{Analysis of Solitons in FZK equation utilizing  Hirota bilinear method} 
	\author{Saugata Dutta\inst{1,a}[0000-0003-3042-1212]\and Prasanta Chatterjee$^*$\inst{1,b}  [0000-0001-9383-6066] \and Snehalata Nasipuri\inst{1,c} [0009-0007-5435-6458]
}
	\authorrunning{S. Dutta et al.}
	\tocauthor{Saugata Dutta, Snehalata Nasipuri, Prasanta Chatterjee}
	\institute{Department of Mathematics, Siksha Bhavana, Visva-Bharati,
		Santiniketan-731235, India,\\
		\email{a.saugatadutta.apd@gmail.com},\\
\email{b.prasantacvb@gmail.com} ($^*$corresponding author)\\
\email{c.snehalatanasipuri@gmail.com}
	}
	
	\maketitle              
	
	\begin{abstract}
		The influence of fractional order parameter $(\alpha)$ in non-linear waves is examined in the fractional Zakharov–Kuznetsov (FZK) equation with the Hirota bilinear approach. Symbolic computation is used for all mathematical calculations. A significant impact of the fractional order parameter is found on the single and multi-soliton solutions. The fact that the structural change is noticeable when $\alpha$ is raised, is crucial to our investigation.
		\keywords{Zakharov–Kuznetsov equation $\cdot$ Fractional order parameter $\cdot$ Hirota bilinear $\cdot$ Computation $\cdot$ Soliton structures.}
	\end{abstract}

	\section{Introduction:}
	In the past decade researchers and scientists of various fields have shown a strong interest in analyzing and evaluating the Partial differential equations with partial and fully integrable nonlinearities  \cite{Chatterjee:2022,Yang:2001}. The nonlinear evolution equations play a significant role in studying nonlinear physical systems. The nonlinear dynamic phenomena associated with the improvements in solitary waves is highly exciting for research. In 1895, Korteweg and de Vries  \cite{KdV:1895} developed a classical and vital equation, i.e. the well-known KdV equation, which reflects nonlinear waves that have unidirectional propagation. The KdV equation governs the interaction of two unidirectional solitary waves and is often researched in applied physics, fibre optics, coastal engineering, etc. Zabusky and Kruskal \cite{Zabusky:1965} numerically investigated this problem for the first time in 1965. They found that the two-solitons progressively restored their original waveforms through collision. The larger (faster) soliton gets a forward phase changes, whereas the smaller (slower) soliton has a backward phase change. An analytical solution to the KdV equation's initial-value issue was presented by Gardner et al. (1967) \cite{Gardner:1967}, particularly for the motion of soliton. In 1971 Hirota \cite{Hirota:1971} also gave the exact solution of the KdV equation for multiple-soliton collisions.  One of the most investigated standard two special dimensional extensions of the KdV equation is the Zakharov-Kuznetsov (ZK) equation \cite{Zakharov:1974}. KdV waves and soliton in a strong, uniform magnetic field can be described by ZK equation where space stretching is isotropic. For the ZK equation, a traveling wave analysis is obtained in \cite{Kourakis:2009}. The enhanced modified extended tanh-function approach is used to develop soliton solutions \cite{Kourakis:2009}. Singular soliton, apparently rigid \cite{Infeld:1987}, periodic solutions \cite{He:2005}, and N-soliton solutions \cite{Q:2011} are investigated for ZK equation. The ZK equation has numerous applications in applied mathematics, engineering, and physics \cite{Munro:1999,Elwakil:2011,saini:2014,Kumar:2021}. Several researchers are getting interest in the study of the FZK equation which has significant applications in applied sciences. Different methods, such as variational iteration method (VIM) \cite{Molliq:2009}, homotopy perturbation method (HPM) \cite{Kumar:2014}, the perturbation-iterative algorithm and residual power series approach \cite{Senol:2018}, the new iterative Sumudu transform method (NISTM) \cite{Prakash:2018} and others are used to study FZK equations.\par
 The idea of differentiation is extended to non-integer or fractional order in fractional differential equations \cite{Diethelm:2002,Kilbas:2006,Zhou:2023}, which require derivatives of non-integer order. An effective tool for simulating and examining systems with anomalous diffusion, long-range interactions, and memory effects is fractional calculus. The fractional differential equation can be expressed in general as  $D^\alpha \gamma(\chi)= \varphi\left(\chi,\gamma(\chi),\gamma'(\chi),\cdots\right)$, where $D^\alpha$denotes the fractional derivative of order $\alpha$, and $\varphi\left(\chi,\gamma(\chi),\gamma'(\chi),\cdots\right)$, is a function that, up to a certain order, involves the dependent variable $\gamma$ and its derivatives. In this work, we have used the Hirota bilinear approach to investigate the effect of $\alpha$ on the solitary wave solution of the FZK problem. To the best of our knowledge, the Hirota-Biliar approach has not yet been used to explore to study the soliton in the FZK equation. Generally, the nonlinear partial differential equation(FDE) for the FZK problem looks like this:
\begin{equation}\label{1}
	D^\alpha_\tau \mu+6\mu\mu_\chi+\mu_{\chi\chi\chi}+\mu_{\chi\gamma\gamma}=0.
\end{equation}
Here $D^{\alpha}_\tau\equiv\frac{\partial^\alpha }{\partial \tau^\alpha}$ and the order of the partial derivative is indicated by the subscripts.Here, the wave function is denoted as $\mu=\mu(\chi,\gamma,\tau)$, and the time and space variables are $\tau$, $\gamma$, and $\chi$, respectively. Furthermore, $\alpha \left(0<\alpha<1\right)$ is the fractional order parameter that appears in the FDE and indicates the derivative's strength. If $\alpha=0$, then there is no derivative operating on the function; otherwise, the derivative is the standard derivative for $\alpha=1$. However, the derivative becomes slower if $\alpha$ is between $0$ and $1$; this is comparable to the delay derivative that was introduced in biomathematics. The independent variables for any evolution equation are $\tau$ (time), $\gamma$ and $\chi$ (space). 
 The wave function $\mu(\chi,\gamma,\tau)$ is a commonly used representation of the amplitude or shape of the soliton solution. It is crucial for understanding wave interactions in plasma situations, energy shifting, shallow-water and wave motion in the ocean, and unaltered long-distance data transmission. The FZK equation is essential to the domains of optical fiber, telecommunications, fluid mechanics, and plasma dynamics.\par
The definitions of the Caputo differential operator and the Riemann-Liouville (RL) fractional differential and integral operators were covered in section: \ref{S2}. The section: \ref{S3.1} describes the Hirota bilinear approach. The sections: \ref{S3.2} and \ref{S3.3}, respectively, describe one-soliton and two-soliton solutions. Finally, section: \ref{S4}, contains the discussion and conclusion of this investigation.
\section{Fractioanl Calculus:}\label{S2}
	To define different fractional order derivatives, many methods have been tried \cite{Miller:1993,Butzer:2000,Machado:2011}. Some of them are,
\subsection{Definition-I:}
The fractional differential operator of Riemann-Lioville(R-L) \cite{Hilfer:2000} $D^\alpha$ of order $\alpha$ is explained as
\begin{equation}\label{2}
	D^\alpha \phi(x)=\left\{
	\begin{array}{cc}
		\frac{d^m}{dx^m}\phi(x),~~\alpha=m;&\\
		\frac{1}{\Gamma(m-\alpha)}\frac{d^m}{dx^m}\int_{0}^{x}\frac{\phi(t)}{(x-t)^{\alpha-m+1}}dt,&~m-1<\alpha<m;
	\end{array}
	\right.
\end{equation}
where $m\in \mathbb{Z}^+, \alpha\in \mathbb{R}^+$ and
\begin{align}\label{3}
	D^{-\alpha}\phi(x)=\frac{1}{\Gamma(\alpha)}\int_{0}^{x}(x-t)^{\alpha-1}\phi(t)dt,0<\alpha\leq1.
\end{align}
\subsection{Definition-II:}The R-L fractional integral operator \cite{Samko:1993} $J^\alpha$ is described as
\begin{align}\label{4}
	J^\alpha\phi(x)=\frac{1}{\Gamma(\alpha)}\int_{0}^{x}(x-t)^{\alpha-1}\phi(t)dt,~t>0,\alpha>0;
\end{align}
and we have,
\begin{align}
	J^\alpha t^\alpha&=\frac{\Gamma(m+1)}{\Gamma(m+\alpha+1)}t^{m+\alpha}\label{5},\\
	D^\alpha t^m&=\frac{\Gamma(m+1)}{\Gamma(m-\alpha+1)}t^{m-\alpha}\label{6}.
\end{align}
\subsection{Definition-III:} The Caputo fractional order differential operator \cite{Almeida:2017} $^cD^\alpha$ of order $\alpha$ is described as
\begin{equation}\label{7}
	^cD^\alpha=\left\{
	\begin{array}{cc}
		&\frac{d^m}{dt^m}\phi(t),~\alpha=m;\\
		&\frac{1}{\Gamma(m-\alpha)}\int_{0}^{x}\frac{\phi^m(t)}{(x-t)^{\alpha-m+1}}dt,~m-1<\alpha<m.
	\end{array}
	\right.
\end{equation}
 \section{Mathematical Construction:}
	Let us take $\xi=\frac{\tau^\alpha}{\Gamma (1+\alpha)}$. Then, we have
\begin{align}\label{8}
\mu(\chi,\gamma,\tau)&=u(x,y,\xi).\nonumber\\
\textnormal{So,}~D^{\alpha}_\tau \mu(\chi,\gamma,\tau)&=D^\alpha_\tau u(x,y,\xi);\nonumber\\
	\textnormal{or}, D^{\alpha}_\tau \mu(\chi,\gamma,\tau)&=\frac{\partial u}{\partial \xi}\cdot D^{\alpha}_\tau(\xi)=\frac{\partial u}{\partial \xi}\cdot 1=\frac{\partial u}{\partial \xi}\equiv u_\xi.
\end{align}
Then, eq. (\ref{1}) can be transformed as
\begin{equation}\label{9}
u_\xi+6uu_x+u_{xxx}+u_{xyy}=0,
\end{equation}
where the order of the partial derivatives is indicated by the subscripts and $u$ is a function of $x$, $y$, and $\xi$.
\subsection{Hirota bilinear method:}\label{S3.1}
A strong and useful technique for investigating soliton solutions of several significant equations is the Hirota bilinear approach \cite{Hirota:1981}. Now, we obtain the bilinear form of eq. (\ref{1}), using the following transformation.
\begin{align}\label{10}
	u=36(\ln f)_{xx}-12(\ln f)_{xy}.
\end{align}
A real function $f = f(x,y, \xi)$, with the partial derivatives with respect to $x$ and $y$ represented by the subscripts. We may derive the bilinear form of eq. (\ref{9}) by substituting eq. (\ref{10}) into eq. (\ref{9}) and using the properties of Hirota bilinear operators.
\begin{align}\label{11}
\left(D_\xi D_x + D_x^4 + D_x^2 D_y^2\right)(f \cdot f)=0,
\end{align}
where $D_x$, $D_y$ and $D_\xi$ are the operators of Hirota bilinear which is defined by
\begin{align}\label{12}
	D_a^mD_b^n(g\cdot h)=\left(\frac{\partial}{\partial a_{1}}-\frac{\partial}{\partial a_{2}}\right)^m\left(\frac{\partial}{\partial b_{1}}-\frac{\partial}{\partial b_{2}}\right)^n\left\{g(a_{1},b_{1})\cdot h(a_{2},b_{2})\right\}|_{a_{1}=a_{2}=a,b_{1}=b_{2}=b}.
\end{align}
Using the operator $F(D)=\left(D_\xi D_x + D_x^4 + D_x^2 D_y^2\right)$,  eq. (\ref{11}) becomes $F(D)\left(f.f\right)=0.$\vspace*{0.15cm}\\
Now, we use the perturbation $f=1+\epsilon f_1+\epsilon^2f_2+\cdots$, where $\epsilon$ is the size of the perturbation.\vspace*{0.15cm}\\
Then, $f\cdot f=(1\cdot 1)+\epsilon(f_1\cdot 1+1 \cdot f_1)+\epsilon^2(f_2 \cdot 1+f_1\cdot f_1+1\cdot f_2)+\cdots$\hspace*{0.25cm},\vspace*{0.15cm}\\
and thus,\\ $F(D)\left(f\cdot f\right)=F(D)(1\cdot 1)+\epsilon F(D)(f_1\cdot 1+1 \cdot f_1)+\epsilon^2 F(D)(f_2\cdot 1+f_1\cdot f_1+1\cdot f_2)+\cdots$
	\subsection{One-soliton solution:}\label{S3.2}
	For the solution of one-soliton,  we assume 
	\begin{align}\label{13}
f(x,\xi)=1+\exp(\eta),~~\eta=kx-ky-k^3\xi+\theta;~~ \xi=\frac{t^\alpha}{\Gamma (1+\alpha)}.
	\end{align}
The one-soliton solution is given by
	\begin{align}\label{14}
		u(x,y,t)=60\frac{k^2}{4}sech^2(\eta/2),
	\end{align}
where $\eta=kx-ky-k^3\xi+\theta$
and $\xi=\frac{t^\alpha}{\Gamma (1+\alpha)}$.
\clearpage
\begin{figure}[h]
	\begin{subfigure}{.31\textwidth}
		\centering
\includegraphics[width=1.2\linewidth]{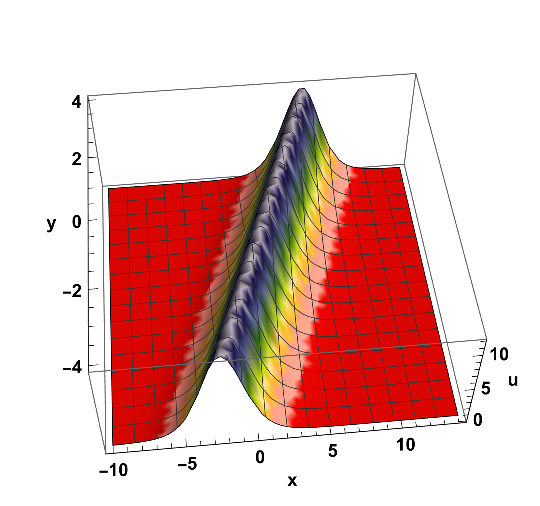}
		\caption{ $\alpha=0.3$ }
		\label{fig:1a}
	\end{subfigure}%
	\hspace*{0.25cm}
	\begin{subfigure}{.31\textwidth}
		\centering
\includegraphics[width=1.2\linewidth]{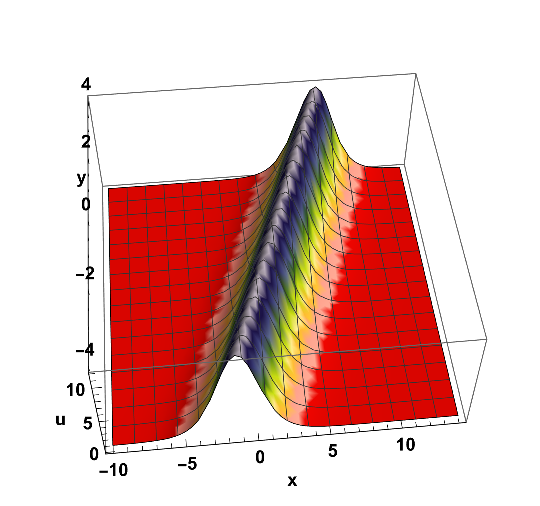}
		\caption{ $\alpha=0.6$}
		\label{fig:1b}
	\end{subfigure}%
	\hspace*{0.25cm}
	\begin{subfigure}{.31\textwidth}
		\centering
\includegraphics[width=1.2\linewidth]{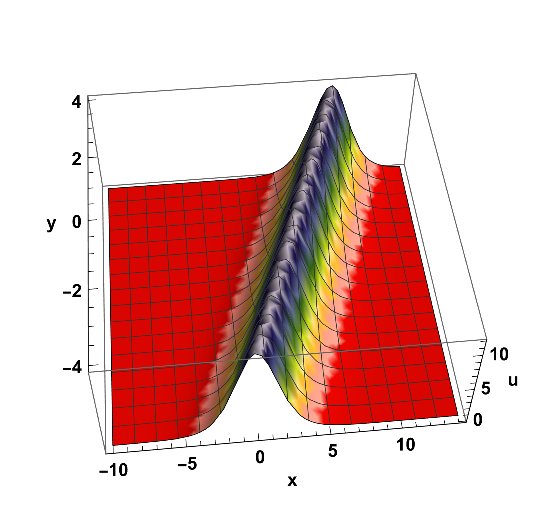}
		\caption{ $\alpha=0.9$}
		\label{fig:1c}
	\end{subfigure}\vspace*{0.5cm}
	\begin{subfigure}{.31\textwidth}
		\centering
\includegraphics[width=1\linewidth]{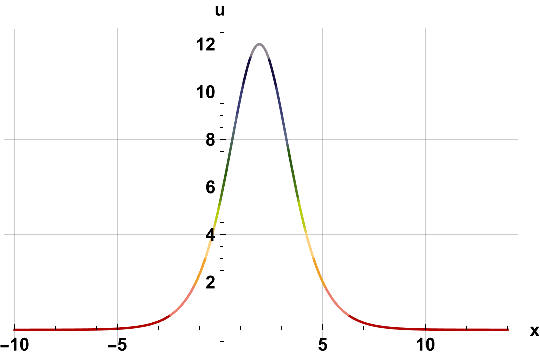}
\caption{$\alpha=0.01$ }
		\label{fig:1d}
	\end{subfigure}%
		\hspace*{0.25cm}
	\begin{subfigure}{.31\textwidth}
		\centering
\includegraphics[width=1\linewidth]{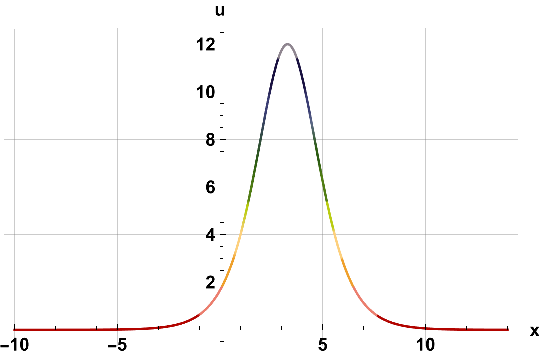}
	\caption{$\alpha=0.5$ }
		\label{fig:1e}
	\end{subfigure}%
		\hspace*{0.25cm}
	\begin{subfigure}{.31\textwidth}
		\centering
\includegraphics[width=\linewidth]{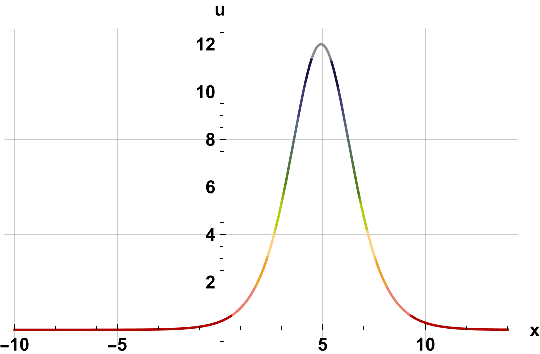}
		\caption{ $\alpha=0.9$}
		\label{fig:1f}
	\end{subfigure}%
	\caption{Structures in three dimension for one-soliton  for (a): $\alpha=0.3$, (b): $\alpha=0.6$, (c): $\alpha=0.9$, and the associated two-dimensional structures for (d): $\alpha=0.01$, (e): $\alpha=0.5$, (f): $\alpha=0.9$ at $t=4.5$ with $k=1,\theta=0.1$.}
	\label{fig:1}
\end{figure}
One-soliton solution of eq. (\ref{1}) is represented in Fig. (\ref{fig:1}) for various values of $\alpha$~(= 0.1, 0.5, and 0.9) with $k=1,\theta=0.1$. The corresponding two-dimensional plots at $t=4.5$ are specified in Figs. (\ref{fig:1d}), (\ref{fig:1e}), and (\ref{fig:1f}), while the three-dimensional plots are shown in Figs. (\ref{fig:1a}), (\ref{fig:1b}), and (\ref{fig:1c}).  As $\alpha$ increases, it is noticed that the solitary soliton's form is distorted, but its amplitude is unchanged. Additionally, a considerable phase change of the solitary wave is observed with increasing strength of the fractional derivative.
\subsection{Two-soliton solution:}\label{S3.3}
	For the solution of two-soliton, we take
	\begin{align}\label{12}
		f(x,\xi)=1+\exp(\eta_1)+\exp(\eta_2)+A_{12}\exp(\eta_1+\eta_2),
	\end{align}
	where $\eta_i=k_ix-k_iy-k_i^3\xi+\theta_i$; $i=1,2$, $A^2_{12}=\left(\frac{k_1-k_2}{k_1+k_2}\right)$ and $\xi=\frac{t^\alpha}{\Gamma (1+\alpha)}$ .\\
	The two-soliton solution is given by
	\begin{align}\label{14}
		u(x,y,t)=A_1sech^2\left(\frac{\eta_1}{2}+ \log|A_{12}|\right)+A_2sech^2\left(\frac{\eta_1}{2}+ \log|A_{12}|\right)
	\end{align}
	where $\eta=k_ix-k_iy-k_i^3\xi+\theta_i~;$ $A_i=\frac{1}{2}k^2_i$ with $i=1,2$ and $\xi=\frac{t^\alpha}{\Gamma (1+\alpha)}$, where $A^2_{12}=\left(\frac{k_1-k_2}{k_1+k_2}\right)$ represents phase shift.
\begin{figure}[h]
	\begin{subfigure}{.33\textwidth}
		\centering
		\includegraphics[width=\textwidth]{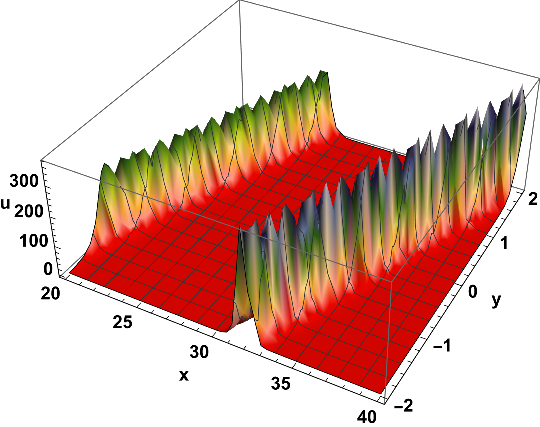}
		\caption{ $\alpha=0.3$}
		\label{fig:2a}
	\end{subfigure}
	\begin{subfigure}{.33\textwidth}
		\centering
		\includegraphics[width=\textwidth]{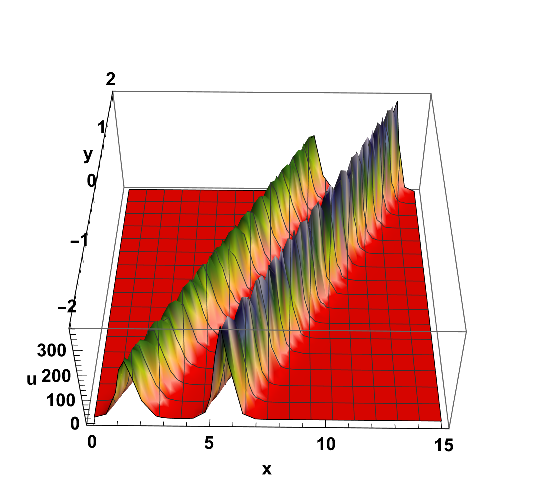}
		\caption{$\alpha=0.6$}
		\label{fig:2b}
	\end{subfigure}
	\begin{subfigure}{.33\textwidth}
		\centering
		\includegraphics[width=\textwidth]{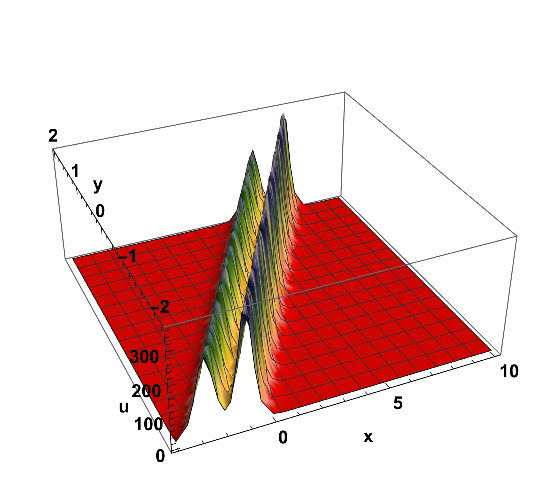}
		\caption{$\alpha=0.9$}
		\label{fig:2c}
	\end{subfigure}
	\caption{Structures in three dimensions for the two-soliton solutions for (a): $\alpha=0.3$, (b): $\alpha=0.6$, and (c): $\alpha=0.9$ with $k_1=1, k_2=2,\theta_1=0.1 ~\text{and}~ \theta_2=0.2$.}
	\label{fig:2}
\end{figure}\\
The three-dimensional structures for the two-soliton solution of eq. (\ref{1}) are shown in Fig. (\ref{fig:2}).  The three-dimensional plots for $\alpha=0.3,0.6$, and $0.9$, respectively, are displayed in Figs. (\ref{fig:2a}), (\ref{fig:2b}), and (\ref{fig:2c}). The amplification of $\alpha$ is shown to cause a noticeable change in the form of the solitary waves. The contour plot indicates that one of the solitons has greater amplitude, and for $\alpha=0.6$, it is seen that the solitary waves are relatively stable.  As $\alpha$ increases, the two-soliton intervening distance gets smaller. Interestingly, for high $\alpha$, it turns out that the intervening gap has grown important with the small period.
\begin{figure}[h!]
		\begin{subfigure}{.33\textwidth}
			\centering
			\includegraphics[width=.8\linewidth]{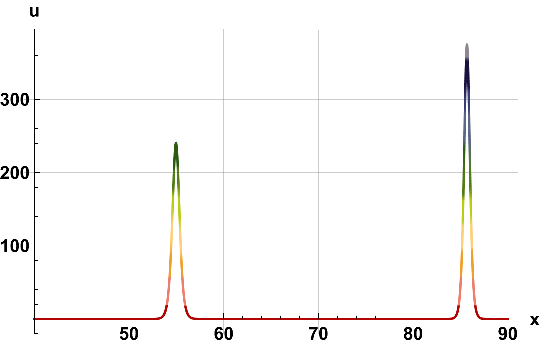}
			\caption{$\alpha=0.1$ }
			\label{fig:sfig1}
		\end{subfigure}%
		\begin{subfigure}{.33\textwidth}
			\centering
	\includegraphics[width=.8\linewidth]{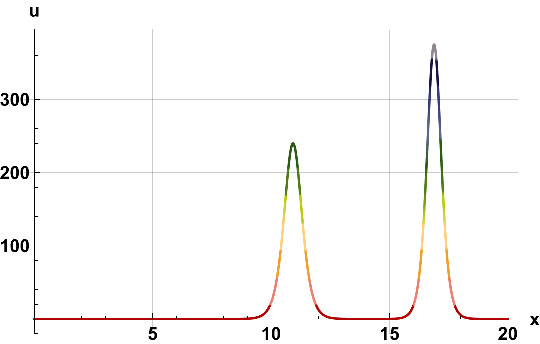}
			\caption{$\alpha=0.3$}
			\label{fig:sfig2}
		\end{subfigure}%
		\begin{subfigure}{.33\textwidth}
			\centering
			\includegraphics[width=.8\linewidth]{ZK/2SS/2D,a,0.5.eps}
			\caption{$\alpha=0.5$}
			\label{fig:sfig3}
		\end{subfigure}
		\begin{subfigure}{.33\textwidth}
			\centering
			\includegraphics[width=.8\linewidth]{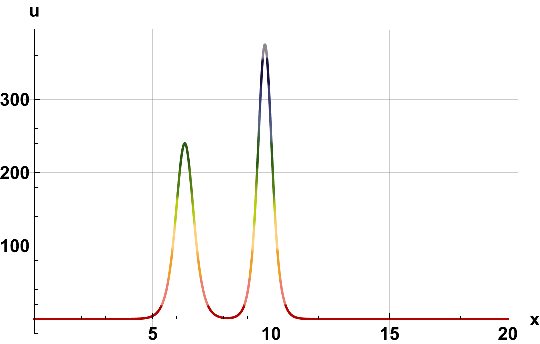}
			\caption{ $\alpha=0.65$ }
			\label{fig:sfig1}
		\end{subfigure}%
  \begin{subfigure}{.33\textwidth}
			\centering
			\includegraphics[width=.8\linewidth]{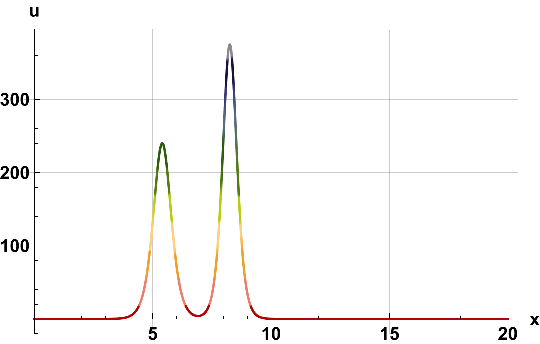}
			\caption{ $\alpha=0.7$ }
			\label{fig:sfig1}
		\end{subfigure}%
  \begin{subfigure}{.33\textwidth}
			\centering
			\includegraphics[width=.8\linewidth]{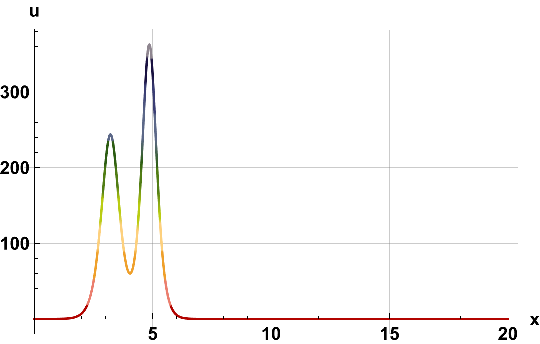}
			\caption{ $\alpha=0.9$ }
			\label{fig:sfig1}
		\end{subfigure}
		\caption{Structures in two dimensions for a two-soliton solution for (a): $\alpha=0.1$, (b): $\alpha=0.3$, (c): $\alpha=0.5$, (d): $\alpha=0.65$, (e): $\alpha=0.7$; (f): $\alpha=0.9$ with $k_1=1, k_2=2,\theta_1=0.1 ~\text{and}~ \theta_2=0.2$.}
		\label{fig:3}
	\end{figure}\\
 The two-dimensional plot of eq. (\ref{1}) for different values of $\alpha$ is shown in Fig. (\ref{fig:3}) when the other fixed parameters are $k_1=1, k_2=2,\theta_1=0.1 ~\text{and}~ \theta_2=0.2$. As $\alpha$ rises, it is seen that both the solitary waves and their inclination to collide rise. It is important to note that soliton interaction is observed when $\alpha$ is moving toward $1$, particularly when $\alpha=0.9$.
	\section{Discussion and conclusion:}\label{S4}
	Within the framework of the FZK equation, the current investigation discusses both one-soliton and two-soliton solutions.  The converted version of the FZK equation is obtained by applying the transformation $\xi=\frac{t^\alpha}{\Gamma (1+\alpha)}$. The bilinear version of the problem is obtained using the Hirota bilinear approach. The one-soliton and two-soliton solutions are also derived via the perturbation approach. A single-soliton solution exhibits a distorted bell-shaped solitary wave. The alteration in $\alpha$ causes a considerable change in the amplitude of the two-soliton solution. significant modifications to soliton structures seen in two-soliton solutions. There is interaction between the two isolated waves when $\alpha$ tends to one. The delicate phase and amplitude changes of solitary waves are observed with the change of the fractional order parameter. This observation greatly aids in the understanding of the fundamentals of many physical issues in the domains of viscoelasticity, electromagnetism,  electrochemistry, signal processing, fluid mechanics, and so on.
	\section*{Acknowledgments:}
 The author, Saugata Dutta (NTA Ref. No.211610066362), expresses sincere gratitude to the University Grants Commission (U.G.C.) of India for providing financial support that has enabled him to conduct this inquiry.
	%
	%
	
\end{document}